\documentclass[11pt]{article}
\usepackage{graphicx}
\def \d {{\rm d}}

\begin{document}

\title{A disintegrating cosmic string}

\author{J. B. Griffiths\thanks{E--mail: {\tt J.B.Griffiths@Lboro.ac.uk}} \ and P.
Docherty  \\ \\ 
Department of Mathematical Sciences, Loughborough University, \\ 
Loughborough, Leics. LE11 3TU, U.K. \\ }

\date{\today}
\maketitle

\begin{abstract}
\noindent
We present a simple sandwich gravitational wave of the Robinson--Trautman family.
This is interpreted as representing a shock wave with a spherical wavefront which
propagates into a Minkowski background minus a wedge. (i.e. the background
contains a cosmic string.) The deficit angle (the tension) of the string
decreases through the gravitational wave, which then ceases. This leaves an
expanding spherical region of Minkowski space behind it. The decay of the cosmic
string over a finite interval of retarded time may be considered to generate the
gravitational wave. 
\end{abstract}

\section{Introduction}

An exact solution of Einstein's equations is known
\cite{GlePul89}--\cite{NutPen92} which describes a snapping cosmic string. The
string is represented by Minkowski space minus a wedge, the deficit angle
representing a string of constant tension. At some event, the string snaps and the
two ends move apart at the speed of light. This generates an impulsive spherical
gravitational wave, inside which the Minkowski space is complete (i.e. there is
no string).

It has recently been argued \cite{PodGri99a} that this solution may be
considered as the impulsive limit of the Robinson--Trautman family of type~N
solutions \cite{RobTra60} in spite of the fact that, in a convenient coordinate
system, the metric in this impulsive limit includes the product of two delta
distributions. This interpretation is here confirmed by investigating the
physical interpretation of a sandwich wave constructed from a particular member
of this family of solutions.

Specifically, we present a particular Robinson--Trautman solution in which the
gravitational wave is non-zero only for a finite interval. It is well known that
these waves have spherical wavefronts. Behind the wave, in this case, is an
expanding spherical region of Minkowski space. However, ahead of the spherical
shock front, the Minkowski space contains a cosmic string. In the
Robinson--Trautman interval, there is also a topological singularity in which the
initial deficit angle reduces uniformly to zero through the wave. This solution
is therefore interpreted as describing a cosmic string which, rather than
snapping, disintegrates to zero in a finite time, and in which the regions of
disintegration propagate along the string with the speed of light.

\section{The Robinson--Trautman type~N solutions}

The Robinson--Trautman solutions of Petrov type N (which are necessarily vacuum)
with zero cosmological constant are given by the line element 
 \begin{equation}
 \d s^2 =2\d u\,\d r
+\Big[K(u)-2r(\log P)_u\Big]\d u^2
-2{r^2\over P^2}\,\d\zeta\,\d\bar\zeta. 
 \label{RTmetric}
 \end{equation} 
 where $K(u)$ can be interpreted as the Gaussian curvature of the 2-surfaces
$2P^{-2}\d\zeta\d\bar\zeta$, and $P(u,\zeta,\bar\zeta)$ is a solution of the
equation 
 \begin{equation}
 2P^2\partial_\zeta\partial_{\bar\zeta}(\log P)=K(u).
 \label{RTeqn}
 \end{equation}
 A general solution of this equation can be expressed as 
 \begin{equation}
 P=\big(1+{\textstyle{1\over2}}KF\bar F\big) 
\big(F_\zeta\bar F_{\bar\zeta}\big)^{-1/2}, 
 \end{equation} 
 where $F=F(u,\zeta)$ is an arbitrary complex function of $u$ and $\zeta$,
holomorphic in $\zeta$. Also, it is always possible to use a coordinate
transformation to put \ $K=-1,0,+1$. \ Finally, it can be shown that, using a
natural tetrad, the only remaining component of the Weyl tensor is given by 
 \begin{equation}
 \bar\Psi_4 =-{1\over r}\left[ P^2\big(\log P\big)_{u\zeta} \right]_{\zeta}
 ={P^2\over2r}F_\zeta 
\left[{1\over F_\zeta}\big(\log F_\zeta\big)_{u\zeta}\right]_\zeta. 
 \end{equation} 
 These solutions always contain singular points on each wave surface, forming
singular lines, but these are generally not well understood.

We will concentrate here on the case in which \ $K=1$, \ and first consider the
Minkowski background. In this case, we can put \ $F=\zeta$, \ which is then a
stereographic coordinate. Putting
\ $\zeta=\sqrt2\cot{\theta\over2}e^{i\phi}$ \ and \ $u=t-r$, \ the metric
(\ref{RTmetric}) takes the form 
 \begin{equation}
 \d s^2 =\d t^2 -\d r^2
-r^2(\d\theta^2+\sin^2\theta\>\d\phi^2), 
 \label{MinkSP}
 \end{equation} 
 from which it is clear that the wave surfaces $u=$ const. are expanding
concentric spheres.

Let us also consider a family of solutions in which \ $F=\zeta^{c+au}$, \ where
$c$ and $a$ are constants. This represents an expanding gravitational wave with
amplitude proportional to $a$. On any wave surface \ $u=$~const., \
$Z=\zeta^{c+au}$ represents the stereographic coordinate. However, if \
$\arg\zeta\in[0,2\pi)$, \ $Z$ includes an excess angle of \ $(c+au-1)2\pi$.

\section{A sandwich Robinson--Trautman wave}

Let us now use the above solution with $K=1$ to construct a sandwich gravitational
wave.

\begin{figure}[hpt]
\begin{center} \includegraphics[scale=0.4, trim=5 5 5 -5]{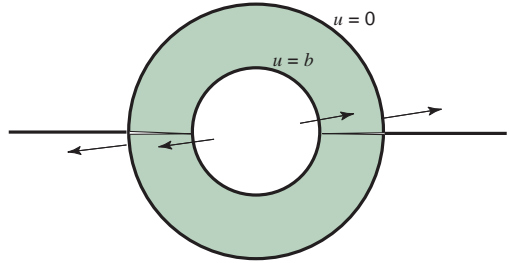}
\caption{ The shaded region represents a sandwich gravitational wave at some time
$t>0$. The expanding spherical wave surfaces are given by $u =$ const. The
interior region behind the wave is Minkowski, while the outer region ahead of the
wave is Minkowski minus a wedge representing a cosmic string. }
\end{center}
\end{figure}

First, let us take the region inside some expanding circle \ $u=b$ \ to be part of
Minkowski space (\ref{MinkSP}). This is a particular case of the
Robinson--Trautman metric (\ref{RTmetric}) with \ $F=\zeta$ \ and \
$\arg\zeta\in[0,2\pi)$. \ Then, let us consider a Robinson--Trautman wave in the
region \ $0\le u\le b$ \ with \ $F=\zeta^{c+au}$, \ as shown in figure~1. In this
case, $b$ represents the duration of the wave. In order for the solution to be
continuous at \ $u=b$, \ it is necessary that \ $c=1-ab$. \ Finally, let us
consider the region ahead of the wave, in which \ $u<0$, \ also to be part of
Minkowski space. For this to be continuous on the wavefront \ $u=0$, \ it is
necessary that this region is represented by the metric (\ref{RTmetric}) with \
$F=\zeta^{1-ab}$. \ According to the above calculation, it can be seen that this
part of Minkowski space has a deficit angle given by $2\pi ab$.

\section{Conclusion}

As described above, we have constructed a very simple sandwich Robinson--Trautman
wave by putting   
 \begin{equation}
 F(u,\zeta) =
 \left\{ \begin{array}{lcl}
 \zeta^{1-ab} &{\rm for} \ u<0 \\
 \noalign{\smallskip}
 \zeta^{1-a(b-u)}
 \qquad &{\rm for} \ 0\le u\le b \\
 \noalign{\smallskip}
 \zeta &{\rm for} \ u>b
\end{array} \right.
 \end{equation}
 with \ $\arg\zeta\in[0,2\pi)$. \ This wave is of type~N for \ $0\le u\le b$ \
with 
 \begin{equation}
 \bar\Psi_4 = -{a\Big(1+{1\over2}(\zeta\bar\zeta)^{1-a(b-u)}\Big)^2 \over
2\Big(1-a(b-u)\Big)\,r\,\zeta^2} (\zeta\bar\zeta)^{a(b-u)}. 
 \end{equation}
 which is unbounded at both poles ($\theta=0,\pi$) representing the
continuations of the strings through the gravitational wave region.

This solution represents a snapping cosmic string in a Minkowski background in
which the tension (deficit angle) of the string reduces uniformly to zero over a
finite interval of retarded time. This decay of the cosmic string may be
considered to generate the gravitational wave.

The above solution is obviously not complete in the sense that the disintegration
of the string does not follow from field equations and initial data. Moreover,
the solution above does not contain its own time reverse, as this would include a
region in which the ingoing and outgoing waves would interact. However, if some
external reason is given for a string to break, the above solution could describe
its subsequent development.

Similar solutions can easily be constructed. For example, $a$ could be negative,
and a string could occur behind the wave rather than in front of it. Also, a
string could be generated and then destroyed, although for this, the energies
associated with the string and the wave would need to be explained. In general
it is known that, for a Robinson--Trautman type~N solution, some singularities
must occur due to the nature of the holomorphic function $F(u,\zeta)$. The above
example shows that these singularities may be associated with string-like
structures. The development or decay of such structures may be considered to
generate the waves, but this should really be associated with a complete theory
for the string.

As well as describing an interesting physical situation, the above solution
appears to give some insight into the character of the singularities that are
known to occur in Robinson--Trautman type~N solutions. Further, since this
solution clearly reduces to the known solution for a snapping string with an
impulsive gravitational wave in the limit as \ $b\to0$, \ while $ab$ remains
finite, this also confirms the interpretation of the impulsive Robinson--Trautman
solutions given previously~\cite{PodGri99a}.

\end{document}